\documentclass[aps,prl,twocolumn,superscriptaddress]{revtex4-1}

\usepackage{times}
\usepackage{mathptmx}
\usepackage{amsmath}
\usepackage{amsfonts}
\usepackage{amssymb}
\usepackage{mathrsfs}
\usepackage{color}
\usepackage[colorlinks]{hyperref}

\usepackage{newtxtext,newtxmath}
\usepackage{pdfrender}
\usepackage[normalem]{ulem}

\newcommand*{\boldgreek}[1]{%
  \textpdfrender{%
    TextRenderingMode=FillStroke,%
    LineWidth=.35pt,%
  }{#1}%
}

\DeclareSymbolFontAlphabet{\mathrsfs}{rsfs}

\definecolor{CiteColor}{rgb}{0,0.5,0}
\hypersetup{citecolor=CiteColor}
\definecolor{RefColor}{rgb}{0.55,0,0}
\hypersetup{linkcolor=RefColor}

\newcommand{\rv}{\bf{\hat{r}}}
\newcommand{\ru}[1]{\hat{r}^{#1}}

\newcommand{\sv}{\bf{\hat{s}}}
\newcommand{\pv}{\hat{\boldgreek{\phi}}}
\newcommand{\pu}[1]{\hat{\phi}^{#1}}

\newcommand{\pAdv}{\phi}
\newcommand{\aM}{\chi}
\newcommand{\ud}{\mathrm{d}}
\newcommand{\ui}{\mathrm{i}}
\newcommand{\beq}{\begin{equation}}
\newcommand{\eeq}{\end{equation}}
\newcommand{\bsubeq}{\begin{subequations}}
\newcommand{\esubeq}{\end{subequations}}

\begin{document}

\title{Spinning Black Holes Fall in Love}

\author{Alexandre Le Tiec}
\affiliation{Laboratoire Univers et Th{\'e}ories, Observatoire de Paris, CNRS, Universit{\'e} PSL, Universit{\'e} de Paris, 92190 Meudon, France}
\affiliation{Centro Brasileiro de Pesquisas F{\'i}sicas (CBPF), Rio de Janeiro, CEP 22290-180, Brazil}

\author{Marc Casals}
\affiliation{Centro Brasileiro de Pesquisas F{\'i}sicas (CBPF), Rio de Janeiro, CEP 22290-180, Brazil}
\affiliation{School of Mathematics and Statistics, University College Dublin, Belfield, Dublin 4, Ireland}

\begin{abstract}
The open question of whether a black hole can become tidally deformed by an external gravitational field has profound implications for fundamental physics, astrophysics and gravitational-wave astronomy. Love tensors characterize the tidal deformability of compact objects such as astrophysical (Kerr) black holes under an external static tidal field. We prove that all Love tensors vanish identically for a Kerr black hole in the nonspinning limit or for an axisymmetric tidal perturbation. In contrast to this result, we show that Love tensors are generically {\it nonzero} for a spinning black hole. Specifically, to linear order in the Kerr black hole spin and the weak perturbing tidal field, we compute in closed form the Love tensors that couple the mass-type and current-type quadrupole moments to the electric-type and magnetic-type quadrupolar tidal fields. For a dimensionless spin $\sim 0.1$, the nonvanishing quadrupolar Love tensors are $\sim 2 \times 10^{-3}$, thus showing that black holes are particularly ``rigid'' compact objects.
\end{abstract}

\maketitle

\textit{Introduction.---}The deformability of a self-gravitating body under the effect of an external tidal field is a question of central interest in gravitational physics. Such tidal deformability can be characterized by a discrete set of \textit{tidal Love numbers} (TLNs)~\cite{PoWi}, the gravitational analogue of the electric susceptibility in electrodynamics. Importantly, the TLNs of a self-gravitating body encode information about its internal structure, such as its composition or equation of state~\cite{Hi.08,Hi.al2.10,Po.al.10}. Those numbers were first introduced by Love~\cite{Lo.09,Lov}, in the context of Newtonian gravitation, to describe the Earth's ocean tides due to its gravitational interaction with the Moon. They now play an important role in understanding the internal structure of the planets of the Solar System~\cite{Ie.al.12,La.16}, and even of exoplanets such as WASP-103b~\cite{Ak.al.19} and the TRAPPIST-1 system~\cite{Au.al.19,Bo.al.20}.

In the context of relativistic gravitation, current and future gravitational-wave measurements of TLNs in binary inspirals provide a novel way of testing the inspiralling compact objects (neutron stars or black holes) and general relativity in the regime of strong gravitational fields \cite{Ca.al.17}. In events that lead to the coalescence of two neutron stars, such as GW170817~\cite{Ab.al3.17} and GW190425~\cite{Ab.al.20}, the tidal effects become important for gravitational-wave frequencies of around 600~Hz, by accelerating the coalescence and affecting the gravitational-wave phase. Those two events have been used to set upper bounds on the tidal deformability of neutron stars, thereby constraining their radii and equation of state at supranuclear densities~\cite{Ab.al2.18,Mo.al.18,De.al.18,Ab.al.20,Ch.20}. Some universal (i.e., equation-of-state independent) I-Love-Q~\cite{YaYu.13} and I-Love-C~\cite{Ch.al.15} relations between the neutron star moment of inertia, quadrupolar TLN, quadrupole moment and compactness can be used to lift degeneracies among parameters in gravitational-wave signals, enhancing the measurability of the tidal effects. Over the coming decades, the observation by the planned LISA mission~\cite{Am.al.17} of the gravitational-wave signals generated by the inspiral of stellar mass compact objects into massive black holes might place constraints on the TLNs of the central body that are roughly eight orders of magnitude more stringent than current ones on neutron stars~\cite{PaMa.19}.

It is widely accepted that all astrophysical black holes are rotating and are thus described by the Kerr family~\cite{Ke.63,Cha} of solutions of the Einstein field equation. Previous works on the tidal deformability of black holes in general relativity have shown that, differently from the Newtonian case, the tidal field can be decomposed into two sectors, according to their parity, often called electric and magnetic, and, importantly, that the TLNs of nonrotating black holes all vanish under a static tidal field \cite{BiPo.09,DaNa2.09,KoSm.12,Ch.al2.13,Gu.15}. This conclusion was extended to slowly rotating black holes, perturbatively in the spin, for a weak and static quadrupolar tidal field: to quadratic order for an axisymmetric quadrupole of electric-type~\cite{Pa.al.15}, and to linear order for a generic quadrupole~\cite{LaPo.15}. Given those remarkable results, there appears to be a widespread expectation that the vanishing of black hole static TLNs extends to a generic rotating Kerr black hole in a generic multipolar tidal environment, e.g. in \cite{Po.16,Po2.16,CaWa.19,Ca.al.19,PaMa.19,CaDu.20,Ta.20,Ch.al.20}. In this Letter we will show that, on the contrary, the static TLNs \footnote{Our definition of TLNs is precisely that introduced in Refs.~\cite{Pa.al.15,Pa.al2.15}. Our use of the term static tidal Love number/tensor corresponds to the fact that a Kerr black hole deforms under a static (i.e, mode frequency $\omega=0$) tidal perturbation, as we explicitly show in \eqref{youpi}, as opposed to nonrotating black holes. It applies regardless of whether the TLNs are real or complex-valued, and whether the corresponding tidal effects are conservative or dissipative, in accord with, e.g., Ref.~[1].} of a Kerr black hole do \textit{not} vanish in general. We show this by fully calculating, for the first time, the induced quadrupole moments on a Kerr black hole due to a static tidal field. The details are given in the companion paper \cite{Le.al.20}.

Throughout this Letter we use units such that $G=c=1$, an overbar denotes the complex conjugation, we use the shorthand $\sum_{\ell m} \equiv \sum_{\ell=2}^\infty \sum_{m=-\ell}^\ell$ where $\ell$ is the multipolar index and $m$ the azimuthal index, as well as the notation $L \equiv i_1 \cdots i_\ell$ for a multi-index made of $\ell$ spatial indices.

\textit{Newtonian Love.---}Consider first an isolated, nonspinning, spherical, Newtonian body of mass $M$ and equilibrium radius $R$. For a weak and slowly varying external tidal field, the induced mass multipole moments $I_L(t)$ are proportional to the applied tidal moments $\mathcal{E}_L(t)$,
\beq\label{lambda}
	I_L = \lambda_\ell \, \mathcal{E}_L \, ,
\eeq
where $\lambda_\ell$ is a constant. For instance, at quadrupolar order the induced mass quadrupole $I_{ij}$ is proportional to the quadrupolar tidal field $\mathcal{E}_{ij}$. This adiabatic approximation holds as long as the typical timescales of the physical processes responsible for adjusting the matter distribution are much shorter than the typical timescale of variation of the tidal environment itself. The tidal deformability parameter $\lambda_\ell$ in \eqref{lambda} depends exclusively on the internal structure of the body and scales as $R^{2\ell+1}$. Introducing the dimensionless TLN $k_\ell$ associated with $\lambda_\ell$, defined via
\beq\label{k}
    k_\ell \equiv - \frac{(2\ell-1)!!}{2(\ell-2)!} \, \frac{\lambda_\ell}{R^{2\ell+1}} \, ,
\eeq
the gravitational potential of the tidally-perturbed Newtonian body can be expanded over spherical harmonics $Y_{\ell m}(\theta,\phi)$ according to \cite{PoWi}
\beq\label{U}
    U = \frac{M}{r} - \sum_{\ell m} \frac{(\ell-2)!}{\ell!} \, \mathcal{E}_{\ell m} \, r^\ell \left[ 1 + 2 k_\ell \left(\frac{R}{r}\right)^{2\ell+1} \right] Y_{\ell m} \, ,
\eeq
where $r$ is the Euclidean distance to the center of mass, and the $2\ell+1$ coefficients $\mathcal{E}_{\ell m}$ are the spherical-harmonic modes of the tidal moment $\mathcal{E}_L$. The growing term $O(r^\ell)$ corresponds to the external $2^\ell$-polar tidal perturbation, and the decaying term $O(r^{-\ell-1})$ to the body's response, proportional to the TLN $k_\ell$.

\textit{Einsteinian Love.---}In general relativity, the tidal environment of a body is fully characterized by \textit{two} families of tidal moments: the electric-type and magnetic-type tidal fields $\mathcal{E}_L$ and $\mathcal{B}_L$. The former are the relativistic analogues of the Newtonian tidal moments introduced above, while the latter have no counterpart in Newtonian gravity. Similarly, the multipolar structure of that body is now characterized by \textit{two} families of multipole moments: the mass-type and current-type multipole moments $M_L$ and $S_L$, which are defined in a coordinate-independent manner for any asymptotically flat, stationary solutions of the vacuum Einstein equation \cite{Ge.70,Ha.74}.

From now on we consider a weak, stationary tidal perturbation of a given compact body. The associated perturbed metric is $\mathring{g}_{\alpha\beta} + h_{\alpha\beta}$, where the background $\mathring{g}_{\alpha\beta}$ is an exact solution of the Einstein equation and the linear metric perturbation can be decomposed according to
\beq
    h_{\alpha\beta} = h^\text{tidal}_{\alpha\beta} + h^\text{resp}_{\alpha\beta} \, .
\eeq
Here, $h^\text{tidal}_{\alpha\beta}$ and $h^\text{resp}_{\alpha\beta}$ are \textit{uniquely} specified as the linearly independent solutions of the linearized Einstein equation in vacuum (outside the body) that have the appropriate asymptotic behavior at large distances. In particular, the growing solution $h^\text{tidal}_{\alpha\beta}$ is unambiguously associated with the perturbing tidal field, while the decaying solution $h^\text{resp}_{\alpha\beta}$ is unambiguously associated with the corresponding linear response of the body (see the Supplemental Material). The perturbed metric $\mathring{g}_{\alpha\beta} + h_{\alpha\beta}^\text{resp}$ is an asymptotically flat, stationary solution of the linearized Einstein equation in vacuum, and the corresponding multipole moments are
\begin{subequations}
    \begin{align}
        M_L &= \mathring{M}_L + \delta M_L \, , \\
        S_L &= \mathring{S}_L + \delta S_L \, ,
    \end{align}
\end{subequations}
where a circle over a quantity indicates it is associated with the background $\mathring{g}_{\alpha\beta}$ and a $\delta$ preceding that it is associated with the linear response $h_{\alpha\beta}^\text{resp}$.

For a nonspinning compact object, the background metric is spherically symmetric. By conservation of parity, the body's linear response contribution to the mass-type (resp.\ current-type) multipole moments can only couple to the electric-type (resp.\ magnetic-type) tidal moments,
\beq\label{elec-mag}
	\delta M_L = \lambda^\text{el}_\ell \, \mathcal{E}_L \quad \text{and} \quad \delta S_L = \lambda^\text{mag}_\ell \, \mathcal{B}_L \, ,
\eeq
where the tidal deformability parameters $\lambda^\text{el}_\ell$ and $\lambda^\text{mag}_\ell$ are constant. If $R$ denotes the areal radius of the central body, then its dimensionless gravito-electric and gravito-magnetic TLNs $k^\text{el}_\ell$ and $k^\text{mag}_\ell$ are defined as per the formula \eqref{k} above. If the central object is spinning, however, then the spherical symmetry of the background metric is broken. Consequently, (i) the multipoles $(\delta M_L,\delta S_L)$ and the tidal moments $(\mathcal{E}_L,\mathcal{B}_L)$ cannot obey simple proportionality relationships akin to Eq.~\eqref{elec-mag}, (ii) the degeneracy of the azimuthal number $m$ is lifted, (iii) fields with different parity can now mix, and (iv) the spherical-harmonic modes $M_{\ell m}$ and $S_{\ell m}$ of $M_L$ and $S_L$ can couple to modes $\mathcal{E}_{\ell'm}$ and $\mathcal{B}_{\ell'm}$ of $\mathcal{E}_L$ and $\mathcal{B}_L$ with $\ell' \neq \ell$ \cite{Le.al.20}.

\textit{Quadrupole moments.---}We consider a Kerr black hole of mass $M$ and spin angular momentum per unit mass $a$ embedded in a weak and stationary, but otherwise completely generic tidal environment. Hence, we work to linear order in the weak tidal perturbation, so that the TLNs are constants. Denoting by $\mathring{M}_{\ell m} = \mathring{M}_\ell \, \delta_{m 0}$ and $\mathring{S}_{\ell m} = \mathring{S}_\ell \, \delta_{m 0}$ the modes of, respectively, the mass-type and current-type multipole moments of the axisymmetric Kerr background spacetime, the multipole moments of the perturbed Kerr geometry read as
\begin{subequations}\label{multipoles_exp}
    \begin{align}
        M_{\ell m} &= \mathring{M}_{\ell m} + \lambda_{\ell m}^{M\mathcal{E}} \mathcal{E}_{\ell m} + \lambda_{\ell m}^{M\mathcal{B}} \mathcal{B}_{\ell m} \, , \\
        S_{\ell m} &= \mathring{S}_{\ell m} + \lambda_{\ell m}^{S\mathcal{E}} \mathcal{E}_{\ell m} + \lambda_{\ell m}^{S\mathcal{B}} \mathcal{B}_{\ell m} \, ,
    \end{align}
\end{subequations}
where $\lambda_{\ell m}^{M\mathcal{E}}$, $\lambda_{\ell m}^{M\mathcal{B}}$, $\lambda_{\ell m}^{S\mathcal{E}}$ and $\lambda_{\ell m}^{S\mathcal{B}}$ are the four families of Kerr TLNs \cite{Pa.al.15,Pa.al2.15}, which are complex-valued. In here, we do not include couplings between different $\ell$ modes because, as we shall show, such couplings are absent in our explicit results for the quadrupole moments $M_{2m}$ and $S_{2m}$. 

We computed (see Supplemental Material) the quadrupolar Kerr TLNs explicitly up to \textit{linear} order in the dimensionless spin parameter $\aM \equiv a/M$. For convenience, we introduce the symbol ``$\doteq$'' for an equality that holds to that order. For any azimuthal number $|m| \leqslant 2$, the result simply reads
\beq\label{TLN_quad}
    \lambda_{2m}^{M\mathcal{E}} \doteq \lambda_{2m}^{S\mathcal{B}} \doteq \frac{\ui m \aM}{180} \, (2M)^5 \quad \text{and} \quad \lambda_{2m}^{M\mathcal{B}} \doteq \lambda_{2m}^{S\mathcal{E}} \doteq 0 \, .
\eeq
The mass-type (resp.\ current-type) quadrupole moment couples \textit{only} to the electric-type (resp.\ magnetic-type) quadrupolar tidal perturbation. The coupling between $(\mathcal{E}_{2m},\mathcal{B}_{2m})$ and $(M_{2m},S_{2m})$ which arises from \eqref{multipoles_exp} and \eqref{TLN_quad} is akin to a Zeeman-like splitting proportional to the azimuthal number $m$ \cite{Pa.al2.15}. Remarkably, the nonvanishing quadrupolar TLNs \eqref{TLN_quad} are purely imaginary. In the related context of {\it nonstatic} tidal perturbations of {\it nonspinning} black holes \cite{Ch.al2.13,Po2.16,Ch2.20}, the imaginary part of the linear response function is well known to give rise to purely dissipative effects, such as tidal heating.

Equations \eqref{multipoles_exp}--\eqref{TLN_quad} imply that a spinning black hole becomes tidally deformed under the effect of a weak, nonaxisymmetric, static tidal field. In particular, while in a binary system, \textit{a spinning black hole falls in Love with its companion.} In the nonspinning limit ($\aM=0$) or for an axisymmetric tidal field ($m=0$), however, the TLNs in \eqref{TLN_quad} vanish, in agreement with the results in \cite{BiPo.09,DaNa2.09,KoSm.12,Ch.al2.13,Gu.15,Pa.al.15}. In fact, we have extended those results to an arbitrarily spinning black hole in a generic multipolar tidal environment, as shown in the Supplemental Material. Our results agree with the previous ones but there is an apparent disagreement with Ref.~\cite{LaPo.15}, who found vanishing black hole TLNs for a generic (nonaxisymmetric) quadrupolar tidal perturbation. This disagreement is a consequence of the use of different splits of the full physical solution into tidal and response contributions. As explained in the Supplemental Material, our tidal/response split relies on the analytic continuation of $\ell \in \mathbb{R}$, which allows us to identify uniquely and unambiguously the two large-radius asymptotic behaviours to be matched onto the known Newtonian solution, whereas the tidal/response split of Ref.~\cite{LaPo.15} relies  on imposing smoothness of the (nonphysical) tidal solution on the black hole horizon.

The event horizon ``radius'' of a slowly spinning Kerr black hole is $2M \left(1+ O(\aM^2)\right)$. Therefore, by analogy with the TLNs \eqref{k} introduced above for a spherical Newtonian body, we define the dimensionless black hole TLNs according to
\beq\label{eq:kME}
    k^{M\mathcal{E}}_{\ell m} \equiv - \frac{(2\ell-1)!!}{2(\ell-2)!} \, \frac{\lambda^{M\mathcal{E}}_{\ell m}}{(2M)^{2\ell+1}} \, ,
\eeq
and similarly for the $M\mathcal{B}$, $S\mathcal{E}$ and $S\mathcal{B}$ couplings. These TLNs generalize to slowly spinning black holes those for nonspinning compact objects \cite{BiPo.09,DaNa2.09}. For a Kerr black hole with spin $\aM \sim 0.1$, Eqs.~\eqref{TLN_quad} and \eqref{eq:kME} imply $|k_{2,\pm2}^{M\mathcal{E}}| = |k_{2,\pm2}^{S\mathcal{B}}| \sim 2 \times 10^{-3}$. This small number could be compared, for instance, to the values $k_2^\text{el} \sim 0.05-0.15$ and $|k_2^\text{mag}| \lesssim 6 \times 10^{-4}$ of the gravito-electric and gravito-magnetic quadrupolar TLNs of a \textit{non}spinning neutron star, depending on the equation of state \cite{BiPo.09,DaNa2.09}. Hence, while spinning black holes do deform like any self-gravitating body, they are particularly ``rigid'' compact objects.

\textit{Tidal Love tensor.---}Having related the spherical-harmonic modes of the quadrupole moments to those of the quadrupolar tidal moments, we now relate $\delta M_{ij}$ and $\delta S_{ij}$ to $\mathcal{E}_{ij}$ and $\mathcal{B}_{ij}$ themselves. Multiplying Eq.~\eqref{multipoles_exp} for $\ell=2$ by $Y_{2m}$, using Eq.~\eqref{TLN_quad}, and summing over modes, we obtain $\ru{i} \ru{j} \delta M_{ij} \propto \ru{i} \pu{j} \mathcal{E}_{ij}$ and $\ru{i} \ru{j} \delta S_{ij} \propto \ru{i} \pu{j} \mathcal{B}_{ij}$, where the unit angular vector $\pv$ is orthogonal to the unit radial vector $\rv$. Therefore, $\delta M_{ij}$ and $\delta S_{ij}$ cannot be simply proportional to $\mathcal{E}_{ij}$ and $\mathcal{B}_{ij}$, respectively. Rather, they must obey more general tensorial relations of the form
\beq\label{M=PE}
    \delta M_{ij} = \sum_{k,l} \lambda_{ijkl} \, \mathcal{E}^{kl} \quad \text{and} \quad \delta S_{ij} = \sum_{k,l} \lambda_{ijkl} \, \mathcal{B}^{kl} \, ,
\eeq
where the constant tensor $\lambda_{ijkl} = O(\aM)$ is the quadrupolar \textit{tidal Love tensor} (TLT) of the Kerr black hole. Such a complication with respect to the nonspinning case \cite{BiPo.09,DaNa2.09} stems from the fact that the black hole spin breaks the spherical symmetry of the background spacetime. Using Eq.~\eqref{TLN_quad}, an explicit calculation shows that the quadrupolar Kerr black hole TLT is given by
\beq\label{TLT}
    \left(\lambda_{ijkl}\right) \doteq \frac{\aM}{180} \; (2M)^5 \;
    \left( \begin{array}{ccc}
        \mathbf{I}_{11} & \mathbf{I}_{12} & \mathbf{I}_{13} \\
        \mathbf{I}_{12} & - \mathbf{I}_{11} & \mathbf{I}_{23} \\
        \mathbf{I}_{13} & \mathbf{I}_{23} & \mathbf{0} \\
    \end{array} \right) \, ,
\eeq
where we introduced the four symmetric and trace-free matrices
\begin{align}\label{matrices}
    \mathbf{I}_{11} &\equiv
        \left( \begin{array}{ccc}
            0 & 1 & 0 \\
            1 & 0 & 0 \\
            0 & 0 & 0 \\
        \end{array} \right) \, , \quad
    \mathbf{I}_{12} \equiv
        \left( \begin{array}{ccc}
            -1 & 0 & 0 \\
            0 & 1 & 0 \\
            0 & 0 & 0 \\
        \end{array} \right) \, , \nonumber \\
    \mathbf{I}_{13} &\equiv 
        \left( \begin{array}{ccc}
            0 & 0 & 0 \\
            0 & 0 & \frac{1}{2} \\
            0 & \frac{1}{2} & 0 \\
        \end{array} \right) \, , \quad
    \mathbf{I}_{23} \equiv 
        \left( \begin{array}{ccc}
            0 & 0 & -\frac{1}{2} \\
            0 & 0 & 0 \\
            -\frac{1}{2} & 0 & 0 \\
        \end{array} \right) \, ,
\end{align}
and with the understanding that the first pair of indices in $\lambda_{ijkl}$ indicates one of these $3 \times 3$ matrices and the second pair refers to an element within it. After rewriting this quadrupolar TLT in a more geometrical form,  the tidally-induced quadrupole moments \eqref{M=PE} of a Kerr black hole explicitly read
\begin{subequations}\label{youpi}
        \begin{align}
            \delta M_{ij} &\doteq \frac{\chi}{90} \, (2M)^5 \sum_{k,l} \mathcal{E}^k_{\phantom{k}(i} \, \varepsilon_{j)kl} \hat{s}^l \, , \\
            \delta S_{ij} &\doteq \frac{\chi}{90} \, (2M)^5
            \sum_{k,l} \mathcal{B}^k_{\phantom{k}(i} \, \varepsilon_{j)kl} \hat{s}^l \, ,
    \end{align}
\end{subequations}
where parentheses around indices denote symmetrization with respect to those indices, $\hat{s}^l$ is a unit vector parallel to the black hole spin, and $\varepsilon_{ijk}$ is the totally antisymmetric Levi-Civita symbol with $\varepsilon_{123} = +1$. The tidally-induced quadrupoles \eqref{youpi} are compatible with the known tidal torquing of a Kerr black hole interacting with a tidal gravitational environment \cite{Le.al.20,Po2.04}.

Let us  consider here the specific but important case that the quadrupolar tidal field $\mathcal{E}_{ij}$ is sourced  by a static particle of mass $\mu \ll M$ a distance $r \gg M$ away from the black hole, in the direction $\rv$. Then, in the Newtonian limit, the formulas \eqref{M=PE}--\eqref{youpi} imply
\beq\label{quadrupole}
    \left(\delta M_{ij}\right) \doteq \frac{\aM}{60} \, (2M)^5 \, \frac{\mu}{r^3} \, \left[ \rv \, \otimes \, (\sv \times \rv) + (\sv\times \rv) \, \otimes \, \rv \right] ,
\eeq
where $\times$ denotes the cross product and $\otimes$ the tensor product. If the particle lies along the axisymmetry axis of the background Kerr geometry (above one of the poles), then $\sv \times \rv = 0$ and $\delta M_{ij}$ vanishes, in agreement with the vanishing of the TLNs \eqref{TLN_quad} for an axisymmetric tidal perturbation. If the particle lies on the equatorial plane, then $(\delta M_{ij}) \propto \rv \otimes \pv + \pv \otimes \rv$, which is the gravitational analogue of the quadrupole moment tensor obtained by setting four electric charges with alternating signs at the corners of a square centered at the origin and whose four sides are tangent to the directions $\rv$ and $\pv$. Interestingly, the purely imaginary TLNs in \eqref{TLN_quad} and the induced mass quadrupole moment \eqref{quadrupole} suggest that the black hole tidal bulge is rotated by $45^\circ$ with respect to the quadrupolar tidal perturbation, which may be interpreted as a ``tidal lag.''

\textit{Speculation.---}As suggested by this tidal lag and as argued in Ref.~\cite{Ch2.20}, the purely imaginary TLNs \eqref{TLN_quad} may give rise to dissipative effects \textit{only}, such as the Kerr tidal torquing discussed in Ref.~\cite{Le.al.20}. However, under the assumption that the induced quadrupole moments \eqref{youpi} also give rise to conservative effects, there is the exciting prospect that the planned space-based gravitational-wave observatory LISA~\cite{Am.al.17} might be able to detect this specific tidal polarization. One of the main sources for LISA is the radiation-reaction driven inspiral of a stellar-mass compact object of mass $\mu$ into a massive black hole of mass $M \gg \mu$. An order-of-magnitude \textit{estimate} of the contribution $\Phi_\text{tidal}$ of the black hole quadrupolar tidal deformability to the total accumulated gravitational-wave phase in such an inspiral is given by applying the formula (11) in Ref.~\cite{PaMa.19}, in which we may tentatively use the typical value $k_1 \sim |k_{22}| \doteq \aM / 60$ derived from \eqref{TLN_quad}--\eqref{eq:kME} for a slowly rotating Kerr black hole \footnote{The definition of the quadrupolar TLN given below Eq.~(2) in Ref.~\cite{PaMa.19} is related to the conventional definition used here by a factor of $(R/M)^5 = 32$ for a (slowly spinning) black hole.}. For instance, for a mass ratio $M/\mu = 10^7$ and a Kerr black hole spin $\aM = 0.1$, this yields $|\Phi_\text{tidal}| \simeq 2 \times 10^3~$rad, much larger than the detectability threshold of $|\Phi_\text{tidal}| > 1~$rad.

\begin{acknowledgments}
We are particularly grateful to Edgardo Franzin for significant contributions to this work. We are grateful to Eric Poisson, Rafael Porto, Adrian Ottewill and Sam Gralla for helpful correspondence, as well as to the whole Capra 23 community for valuable feedback. ALT acknowledges the financial support of the Action F{\'e}d{\'e}ratrice PhyFOG and of the Scientific Council of the Paris Observatory. Both authors acknowledge partial financial support by CNPq (Brazil), process no. 310200/2017-2 and 312917/2019-8.
\end{acknowledgments}

\section*{Supplemental Material}

\textit{Curvature scalar.---}Contracting the Weyl tensor $C_{\alpha\beta\gamma\delta}$ with the vectors $\ell^\alpha$ and $m^\alpha$ of a Newman-Penrose null tetrad~\cite{NePe.62} yields the curvature scalar $\psi_0 \equiv C_{\alpha\beta\gamma\delta} \ell^\alpha m^\beta \ell^\gamma m^\delta$. By considering the relation between the gravitational potential (3) and the Weyl tensor in the Newtonian limit where $c \to \infty$, we find for the Newtonian curvature scalar in the flat spacetime limit of the Kinnersley tetrad~\cite{Ki.69}:
\beq\label{psi0_Newt}
	\lim_{c\to\infty} \psi_0 = \sum_{\ell m} \alpha_{\ell m}(t) \, r^{\ell-2} \left[ 1 + 2 k_\ell \left(\frac{R}{r}\right)^{2\ell+1} \right] {}_{2}Y_{\ell m}(\theta,\phi) \, ,
\eeq
where $\alpha_{\ell m} \equiv [(\ell+2)(\ell+1)/(\ell(\ell-1))]^{1/2} \, \mathcal{E}_{\ell m}$ and ${}_{2}Y_{\ell m}(\theta,\phi)$ is a spin-2 spherical harmonic~\cite{NePe.66}. Similarly to Eq.~(3), the term $O(r^{\ell-2})$ in \eqref{psi0_Newt} corresponds to the external $2^\ell$-polar tidal perturbation, and the term $O(r^{-\ell-3})$ to the body's response. The formula \eqref{psi0_Newt} can easily be generalized to an axisymmetric, spinning Newtonian body, for which $k_\ell \to k_{\ell m}$.

From now on we shall consider a Kerr black hole of mass $M$ and spin angular momentum per unit mass $a$. Using advanced Kerr coordinates $(x^\alpha) = (v,r,\theta,\pAdv)$, the Kerr metric $\mathring{g}_{\alpha\beta}$ is given by~\cite{Ke.63,Cha}
\begin{align}\label{Kerr}
    \mathring{g}_{\alpha\beta} \, \ud x^\alpha \ud x^\beta = &- \left( 1 - \frac{2Mr}{\Sigma} \right) \ud v^2 + 2 \ud v \ud r \nonumber \\
    &- \frac{4 M r}{\Sigma} \, a \sin^2{\theta} \, \ud v \ud \pAdv + \Sigma \, \ud \theta^2 - 2a \sin^2{\theta} \, \ud r \ud \pAdv \nonumber \\
    &+ \left( r^2 + a^2 + \frac{2Mr}{\Sigma} \, a^2 \sin^2{\theta} \right) \sin^2{\theta} \, \ud \pAdv^2 \, ,
\end{align}
where $\Sigma \equiv r^2 + a^2 \cos^2{\theta}$. The coordinate radii of the event ($+$) and Cauchy ($-$) horizons are $r_\pm = M \pm \sqrt{M^2-a^2}$. This black hole is embedded in a weak and slowly varying, but otherwise completely generic tidal environment. Slowly varying means that we can neglect time derivatives, but we emphasize that it is nonstatic as it has a \textit{parametric} dependence on $v$. The tidal field being weak means that we can use the framework of the linearized Einstein equation. This tidal environment is fully characterized by two sets of symmetric and trace-free tensors (angle brackets notation), both defined out of the Weyl tensor $C_{\alpha\beta\gamma\delta}$ and its covariant derivatives (semicolon notation), and evaluated in the black hole's local asymptotic rest frame: the electric-type and magnetic-type tidal fields \cite{Zh.86,BiPo.09,Pa.al.15}
\begin{subequations}\label{EL_BL_GR}
	\begin{align}
		\mathcal{E}_L(v) &\equiv {[(\ell-2)!]}^{-1} \, C_{0 \langle i_1 |0| i_2;i_3 \cdots i_\ell \rangle} (v) \, , \label{E_L_GR} \\
		\mathcal{B}_L(v) &\equiv \frac{3}{2} \, {[(\ell+1)(\ell-2)!]}^{-1} \, \epsilon_{jk\langle i_1} C_{i_2|0jk|;i_3 \cdots i_\ell \rangle}(v) \, , \label{B_L}
	\end{align}
\end{subequations}
where $\epsilon_{ijk}$ is the totally antisymmetric permutation symbol. The electric-type tidal tensors \eqref{E_L_GR} are the general relativistic analogues of the Newtonian tidal tensors. The magnetic-type tidal tensors \eqref{B_L} have no counterpart in Newtonian gravity.

While the curvature Weyl scalar vanishes for the Kerr background \eqref{Kerr}, Teukolsky~\cite{Te.73} managed to derive a decoupled equation for the \textit{perturbed} $\psi_0$. He did so with the choice of the Kinnersley tetrad and showed that, furthermore, the solution admits separation by variables:
\beq\label{psi0}
	\psi_0 = \sum_{\ell m} z_{\ell m}(v) \, R_{\ell m}(r) \, {}_{2}Y_{\ell m}(\theta,\pAdv) \, ,
\eeq
for zero frequency. The coefficients $z_{\ell m} \!=\! \alpha_{\ell m} \!+\! \ui \beta_{\ell m}$ are slowly varying functions of time $v$, where $\alpha_{\ell m}$ and $\beta_{\ell m}$ are related to the spherical-harmonic modes $\mathcal{E}_{\ell m}$ and $\mathcal{B}_{\ell m}$ of the electric-type and magnetic-type tidal moments \eqref{EL_BL_GR} \cite{Ch.al.13}. For instance, at the leading quadrupolar order $\alpha_{2m} = \sqrt{6} \, \mathcal{E}_{2m}$ and $\beta_{2m} = \sqrt{6} \, \mathcal{B}_{2m}$. The radial factor $R_{\ell m}$ in the mode decomposition \eqref{psi0} satisfies the zero-frequency radial Teukolsky equation~\cite{TePr.74}
\begin{align}\label{Teuk}
    &x(x+1) R''_{\ell m}(x) + (6x + 3 + 2 \ui m \gamma) R'_{\ell m}(x) \nonumber \\
    &+ \biggl[ 4\ui m \gamma \, \frac{2x+1}{x(x+1)} - (\ell+3)(\ell-2) \biggr] R_{\ell m}(x) = 0 \, ,
\end{align}
where $x \equiv (r - r_+) / (r_+ - r_-)$ and $\gamma \equiv a / (r_+ - r_-)$. While the physical regime requires $\ell\in \mathbb{L} \equiv \mathbb{Z}^+ \setminus \{1\}$, we shall temporarily carry out the calculations for a {\it generic} $\ell\in\mathbb{R}$, for the following reason. In the Newtonian limit, the identification of the growing tidal perturbation and the associated decaying body's linear response in the gravitational potential $U$ in (3) is unambiguous. This is thanks to the linearity of the Poisson equation satisfied by $U$ and the Euclidean nature of 3-space, which provides a unique and unambiguous notion of distance. In general relativity, however, the identification of the tidal and response contributions is quite subtle. In particular, Gralla \cite{Gr.18} pointed out that the value of the TLNs depends on the choice of radial coordinate. Moreover, given a choice of coordinates and thus of response contribution, the tidal contribution remains ambiguous for $\ell\in\mathbb{L}$ because one can always shift a multiple of the decaying (subdominant) response contribution to the growing (dominant) tidal contribution \cite{BiPo.09,Pa.al2.15}. We henceforth consider Eq.~\eqref{Teuk} with $\ell \in\mathbb{R}$, since this allows us to naturally resolve the two ambiguities, as we shall next see.

\textit{Tidal/response solutions.---}We obtain the specific physical solution by requiring that the curvature scalar $\psi_0$ is smooth at the event horizon $r=r_+$  when written in the Hartle-Hawking tetrad~\cite{HaHa.72}, which is smooth there. The corresponding physical radial factor $R_{\ell m}(x)$, which is a solution of Eq.~\eqref{Teuk}, is thus uniquely determined (up to a normalization) and can be expressed as
\begin{align}\label{sol_Kerr}
    R_{\ell m}(x) = (r_+-r_-)^{\ell-2} \, \bigl[ \hat{R}_{\ell m}^\text{tidal}(x) + 2k_{\ell m} \hat{R}_{\ell m}^\text{resp}(x) \bigr] \, ,
\end{align}
where
\begin{subequations}\label{tidal-resp}
    \begin{align}
        \hat{R}_{\ell m}^\text{tidal} &\equiv x^{\ell-2} \; \frac{F(-\ell-2,-\ell-2\ui m\gamma,-2\ell;-1/x)}{(1+1/x)^2} \, , \label{R_tidal} \\
        \hat{R}_{\ell m}^\text{resp} &\equiv x^{-\ell-3} \; \frac{F(\ell-1,\ell+1-2\ui m\gamma,2\ell+2;-1/x)}{(1+1/x)^2} \, , \label{R_resp}
    \end{align}
\end{subequations}
with $F(a,b,c;z)$ the Gaussian hypergeometric function and
\beq\label{k_phys}
    k_{\ell m} = - \frac{\ui}{4\pi} \, \sinh{(2 \pi m \gamma)} \, |\Gamma(\ell+1+2\ui m \gamma)|^2 \, \frac{(\ell-2)!(\ell+2)!}{(2\ell)!(2\ell+1)!} \, .
\eeq
Here, $\hat{R}_{\ell m}^\text{tidal}$ and $\hat{R}_{\ell m}^\text{resp}$ are two linearly independent solutions of the radial Teukolsky equation \eqref{Teuk}, which are exchanged under the substitution $\ell \to -\ell-1$. While, as said, the physical $\psi_0$ (constructed from $R_{\ell m}$) in the  Hartle-Hawking tetrad should be---and is---smooth at the Kerr horizon, the contributions to it constructed from $\hat{R}^\text{tidal}_{\ell m}$ and $\hat{R}^\text{resp}_{\ell m}$ are both regular but, for $m\gamma \neq 0$, nonsmooth there. They need not be smooth on the horizon since neither of them is separately physically accessible.

The solutions $\hat{R}_{\ell m}^\text{tidal}$ and $\hat{R}_{\ell m}^\text{resp}$ in the formulas \eqref{sol_Kerr} and \eqref{tidal-resp} are specified uniquely and unambiguously by the fact that they carry separately the two linearly independent  asymptotic behaviours $x^{\ell-2}$ and $x^{-\ell-3}$ as $r\to\infty$ for $\ell \in \mathbb{R}$. Crucially, the  \textit{analytic continuation} to $\ell \in \mathbb{R}$ that we have carried out allows us to ascertain whether or not a given solution which contains the asymptotic behaviour $x^{\ell-2}$ also contains the asymptotic behaviour $x^{-\ell-3}$ within it since, for $\ell \in \mathbb{R}$, neither asymptotic behaviour can be considered to be dominant/subdominant (this is further exemplified below \eqref{Gtidal-Gresp} in the case of the Hertz potential). For $\ell \in \mathbb{L}$, the asymptotic behaviours $x^{\ell-2}$ and $x^{-\ell-3}$ physically correspond to the Newtonian tidal field and body's linear response, respectively.

As pointed out in Refs.~\cite{St.15,Pa.al2.15} and shown in Ref.~\cite{Ch.al2.13}, analytic continuation is a natural method~\footnote{We note that taking $\ell\in\mathbb{R}$ is a  mathematical trick which is not uncommon within black hole perturbation literature, such as in the Appendix of Ref.~\cite{Pa.76}.} to obtain a unique and unambiguous tidal/response split that is the general relativistic extension of the unique and unambiguous Newtonian tidal/response split in the rotating version of (3) and \eqref{psi0_Newt}. More precisely, the radial factor \eqref{sol_Kerr} reads
\beq\label{psi0_schema}
    R_{\ell m} = r^{\ell-2} \left( 1 + \cdots \right) + 2k_{\ell m} (r_+-r_-)^{2\ell+1} r^{-\ell-3} \left( 1 + \cdots \right) ,
\eeq
where the dots denote relativistic corrections to the rotating version of the Newtonian \eqref{psi0_Newt}. We determined the overall normalization of $R_{\ell m}$ so that the coefficient of the asymptotic term $r^{\ell-2}$ as $r\to\infty$ times $z_{\ell m}$ is in agreement with the asymptotic form of $\psi_0$ in terms of the spherical-harmonic modes $\mathcal{E}_{\ell m}$ and $\mathcal{B}_{\ell m}$ of the tidal moments \cite{Ch.al.13}.

Comparing the relativistic formula \eqref{psi0_schema} with the Newtonian formula \eqref{psi0_Newt} with $k_\ell \to k_{\ell m}$, the constant coefficients \eqref{k_phys} can be interpreted as the ``Newtonian TLNs'' of a Kerr black hole. (Notice that for a \textit{spinning}, tidally-perturbed Newtonian body, the TLNs can be shown to be complex-valued in general \cite{Le.al.20}.) In the particular cases of a Schwarzschild black hole ($a=0$) and axisymmetric ($m=0$) tidal perturbations of Kerr, those coefficients are identically zero for $\ell\in \mathbb{L}$, and so the response vanishes, yielding zero relativistic black hole TLNs (9) in these cases. This is in agreement with the results in the literature~\cite{BiPo.09,DaNa2.09,KoSm.12,Ch.al2.13,Gu.15,Pa.al.15}. However, the response $2 k_{\ell m} \hat{R}_{\ell m}^\text{resp}$ is {\it not} zero in general in Kerr. In particular, for a small spin $\aM = a/M$, the quadrupolar coefficients \eqref{k_phys} reduce to $k_{2m} \doteq - \ui m \aM/120$, in agreement with the nonvanishing quadrupolar TLNs (8)--(9).

This result appears to be in disagreement with Ref.~\cite{LaPo.15}, who found vanishing TLNs for a generic (i.e. nonaxisymmetric) quadrupolar tidal perturbation of a slowly spinning black hole, to linear order in spin. This disagreement stems from different tidal/response splits of the full physical solution. As we have written above, when setting $\ell\in \mathbb{L}$, there is no unique split in the physical field between a growing solution (identified as the tidal field contribution) and the decaying solution (identified as the response contribution). To lift this ambiguity, Ref.~\cite{LaPo.15} adopted the prescription of imposing smoothness of the tidal solution everywhere, including on the event horizon. Since the decaying solution \eqref{R_resp} is not smooth on the horizon, the linear combination \eqref{sol_Kerr} is the only (up to a normalization) smooth, growing solution, which, under the prescription of Ref.~\cite{LaPo.15}, is then identified as the tidal contribution, yielding zero response contribution and thus vanishing TLNs. As noted, however, this prescription imposes  a  physical  condition  (smoothness) on a nonphysical field (namely, the tidal field).  Furthermore, such prescription, unlike ours, yields a tidal/response split which,  in the nonrelativistic limit, does not correctly reproduce the Newtonian tidal/response  split in Eq.~\eqref{psi0_Newt}, as observed when taking $\ell\in\mathbb{R}$.

\textit{Hertz potential.---}To derive the response part of the metric perturbation from the Weyl scalar \eqref{psi0} with Eqs.~\eqref{sol_Kerr}--\eqref{k_phys}, we apply the Hertz potential  formalism \cite{Ch.75,CoKe.74,KeCo.79,St.79,Wa.78}. The first step consists in obtaining a Hertz potential $\Psi$ by solving a fourth-order linear partial differential equation with the Weyl scalar $\psi_0$ as the source. This can be done mode by mode \cite{Or.03},
\beq\label{Psibar}
	\bar{\Psi} = \sum_{\ell m} z_{\ell m}(v) \, G_{\ell m}(r) \, {}_{2}Y_{\ell m}(\theta,\pAdv) \, ,
\eeq
where the radial function of the (complex-conjugated) Hertz potential inherits the tidal-response split of the Weyl scalar, according to
\beq\label{G=Gtidal+Gresp}
    G_{\ell m}(x) = g_\ell \, (r_+-r_-)^{\ell+2} \, \bigl[ \hat{G}_{\ell m}^\text{tidal}(x) + 2k_{\ell m} \hat{G}_{\ell m}^\text{resp}(x) \bigr] \, ,
\eeq
with $g_\ell \equiv 2 \Gamma(\ell+3)/(\Gamma(\ell+1) [(\ell+2)(\ell+1)\ell(\ell-1)]^2)$ and
\begin{subequations}\label{Gtidal-Gresp}
    \begin{align}
        & \hat{G}_{\ell m}^\text{tidal} = x^\ell (1+x)^2 \, F(-\ell+2,-\ell-2\ui m \gamma,-2\ell;-1/x) \, , \label{G_tidal} \\
        & \hat{G}_{\ell m}^\text{resp} = \frac{(1+x)^2}{x^{\ell+1}} F(\ell+3,\ell+1-2\ui m \gamma,2\ell+2;-1/x) \, . \label{G_resp}
    \end{align}
\end{subequations}
The decomposition \eqref{G=Gtidal+Gresp} into tidal and response contributions in the (complex-conjugated) Hertz potential inherited from the Weyl scalar in Eq.~\eqref{sol_Kerr} has a delicate structure: As one takes $\ell\in \mathbb{L}$, $G_{\ell m}$ becomes a polynomial in $r$ of order $\ell+2$ and so one might be tempted to conclude that the response vanishes in the physical regime $\ell\in \mathbb{L}$. However,  analytic continuation to $\ell \in \mathbb{R}$ shows that this is a mere illusion since such polynomial behaviour arises from the fact that $\hat{G}_{\ell m}^\text{tidal}$ contains this polynomial plus an infinite series in $1/x$ which exactly cancels out with $2k_{\ell m} \hat{G}_{\ell m}^\text{resp}$ when taking $\ell\in \mathbb{L}$.

\textit{Metric reconstruction.---}Once the response contribution to the Hertz potential is known, one can obtain the corresponding metric perturbation $h^\text{resp}_{\alpha\beta}$ by applying a certain second order differential operator to it, e.g. \cite{Ch.75,LoWh.02,YuGo.06}. We carried out the calculation explicitly up to \textit{linear} order in the dimensionless spin parameter $\aM = a/M$. In the so-called ingoing radiation gauge, specified by the conditions $\ell^\alpha h^\text{resp}_{\alpha\beta} = 0$  and  $\mathring{g}^{\alpha\beta} h^\text{resp}_{\alpha\beta} = 0$, we obtain
\beq\label{h_IRG}
	h^\text{resp}_{\alpha\beta} = \ell_\alpha \ell_\beta h^\text{resp}_{nn} + 2 \Re{\bigl[m_\alpha m_\beta h^\text{resp}_{\bar{m}\bar{m}} - 2 \ell_{(\alpha} m_{\beta)} h^\text{resp}_{n\bar{m}}\bigr]}  \, ,
\eeq
with
\begin{subequations}\label{metric_resp}
	\begin{align}
		h^\text{resp}_{nn} \doteq &- 2\aM \sum_{\ell m} C_\ell \, \mathcal{Q}_\ell \; \Re{\left(\ui m \, \bar{z}_{\ell m} \bar{Y}_{\ell m}\right)} , \\
		h^\text{resp}_{n\bar{m}} \doteq &- \sqrt{2} \aM \sum_{\ell m} \ui m \, C_\ell \, \sqrt{\frac{(\ell-1)!}{(\ell+1)!}} \; r \mathcal{Q}'_\ell \; \bar{z}_{\ell m} \, {}_{1}\bar{Y}_{\ell m} \, , \\
		h^\text{resp}_{\bar{m}\bar{m}} \doteq &- 2\aM \sum_{\ell m} \ui m \, C_\ell \, \sqrt{\frac{(\ell-2)!}{(\ell+2)!}} \left( r^2 \mathcal{Q}''_\ell + 2r \mathcal{Q}'_\ell - 2\mathcal{Q}_\ell \right) \bar{z}_{\ell m} \, {}_{2}\bar{Y}_{\ell m} \, ,
	\end{align}
\end{subequations}
where we used $C_\ell \equiv \ell! [(\ell-2)!]^{3/2} / \bigl( {(2\ell)! [(\ell+2)!]^{1/2}} \bigr) \, (2M)^\ell$ and $\mathcal{Q}_\ell(r) \equiv (r/M-2)^2 Q_\ell''(r/M-1)$, with
$Q_\ell$ the Legendre function of the second kind. The $\ell$ modes of the tetrad components \eqref{metric_resp} scale as $O\bigl(r^{-\ell-1}\bigr)$.

\textit{Multipole moments.---}All our results so far were valid for a tidal field which is slowly varying but is allowed to possess a parametric dependence on time $v$. From now on we neglect the slow parametric evolution of the coefficients $z_{\ell m}(v)$ in the metric perturbation \eqref{metric_resp}. Hence, the metric $g_{\alpha\beta} = \mathring{g}_{\alpha\beta} + h^\text{resp}_{\alpha\beta}$ is a stationary, asymptotically flat, vacuum solution of the linearized Einstein equation, and its Geroch-Hansen \cite{Ge.70,Ha.74} multipole moments are well defined.

More precisely, we compute the norm $\lambda$ and scalar twist $\omega$ associated with the timelike Killing field $\xi^\alpha$ of the perturbed Kerr geometry $g_{\alpha\beta}$, and the 3-metric $\gamma_{\alpha\beta} = \lambda g_{\alpha\beta} + \xi_\alpha \xi_\beta$ on any spacelike hypersurface orthogonal to $\xi^\alpha$. We then perform a conformal transformation of $\gamma_{\alpha\beta}$, with a choice of conformal factor $\Omega$ that ensures that the mass dipole moment vanishes, thus taking the multipole moments about the center-of-mass of the perturbed black hole. The mass-type and current-type quadrupole moments $M_{ij}$ and $S_{ij}$ are given by the formula
\beq\label{quadrupoles}
    M_{ij} + \ui S_{ij} = \frac{1}{3} \left( \tilde{D}_{\langle i} \tilde{D}_{j \rangle} \tilde{\Phi} - \frac{1}{2} \tilde{R}_{\langle ij \rangle} \tilde{\Phi} \right)\bigg|_\Lambda \, ,
\eeq
where the scalar field $\tilde{\Phi} \equiv (\Phi_\text{M} + \ui \Phi_\text{S}) / \sqrt{\Omega}$ is a conformally rescaled linear combination of the mass-type and current-type potentials $\Phi_\text{M} \doteq 1/(4\lambda) - \lambda/4$ and $\Phi_\text{M} = \omega / (2\lambda)$, while $\tilde{D}_\alpha$ is the covariant derivative compatible with the conformal metric $\tilde{\gamma}_{\alpha\beta} = \Omega^2 \gamma_{\alpha\beta}$ and $\tilde{R}_{\alpha\beta}$ is the associated Ricci tensor. The right-hand side of Eq.~\eqref{quadrupoles} is evaluated at  spatial infinity, which corresponds to the point $\Lambda$ of the extended 3-manifold. The results (8) are obtained by taking the spherical-harmonic modes of \eqref{quadrupoles} and comparing to the formulas (7).

\hspace{-0.05cm}\textit{Effective action.---}It is well-known in effective field theory (EFT) that finite-size effects correspond to augmenting the effective (spinning) point particle action by nonminimal worldline couplings to curvature \cite{DaEs.96,GoRo.06}. The identification of the Kerr TLNs or TLT computed here with analogous quantities at the level of an effective action (so-called Wilson coefficients) requires a proper matching calculation. Such a matching was performed in the case of nonspinning black holes \cite{KoSm.12,Ch.al2.13,Po2.16}, but remains to be done for rotating (Kerr) black holes. Notice that vanishing black hole TLNs, in the sense of vanishing Wilson coefficients in the effective action, would be at odds with the naturalness dogma: since there is no apparent enhanced symmetry in the effective action describing black holes with vanishing TLNs, this would imply an unnatural fine-tuning from the EFT point of view \cite{Po2.16,Po.16}.

In addition to the open question of matching our Kerr black hole perturbative calculation to an EFT for a spinning particle with finite-size effects, it will also be necessary to address the influence of black hole tidal deformability in the gravitational-wave emission itself, before concluding to the observability of this effect through the gravitational-wave phase.

\bibliography{}

\end{document}